\documentclass[
    aip,
    jmp,
    preprint,
    amsmath,amssymb,
    aps
]{revtex4-2}
\usepackage[dvipdfmx]{graphicx}
\usepackage{bm,ascmac,amssymb,amsmath}
\usepackage{color}
\usepackage{empheq}

\begin{document}

\title{Optimization of fluid mixing by reinforcement learning \\ using limit cycles of a dynamical system}
\affiliation{Graduate School of Engineering Science, the University of Osaka,\\ 1-3 Machikaneyama, Toyonaka, Osaka, 560-8531 Japan}
\author{Masaki Shimizu}
\email{shimizu@me.es.osaka-u.ac.jp}
\author{Daiki Watanabe}
\author{Takumi Aoyagi}
\author{Susumu Goto}
\date{\today}

\begin{abstract}
We propose a method to overcome the difficulties encountered when applying reinforcement learning to fluid mixing processes. The proposed method has two main features: (i) it does not require detailed measurements of the flow state, and (ii) by effectively exploiting a stable limit cycle of a two-dimensional dynamical system (the Li\'enard system), it can stably perform optimization without imposing explicit constraints on the control parameters. As an illustrative example, we optimize a process in which a fluid contained in a cylindrical vessel is mixed by periodically rotating the vessel. The resulting optimal vessel motion is physically reasonable: it reverses its direction of rotation before a solid-body rotation state is established. Furthermore, even when the fluid viscosity increases with time during the mixing process, the method can continuously adapt the control parameters to the changing viscosity.
\end{abstract}

\maketitle

\section{Introduction}
\label{sec:intro}

Machine learning~\cite{bishop2006pattern} has become a powerful tool in science and engineering, with applications spanning a wide range of disciplines. Reinforcement learning~\cite{sutton1998reinforcement}, particularly deep reinforcement learning, has achieved breakthroughs in board games such as Go and chess~\cite{silver2018general} and has also been applied to autonomous driving and robotic control. More recently, it has become a key technology in the training process of large language models (LLMs)~\cite{yu2026dapo}. In this study, we consider the application of reinforcement learning to flow control. Reinforcement learning for flow control has attracted considerable attention since the drag-reduction study by Rabault et al.~\cite{rabault2019artificial}, and various methods have been developed to improve the efficiency of the enormous amount of training required~\cite{zheng2022data}. Beyond direct flow control, reinforcement learning has also been applied to agent navigation in thermally and chemically driven environments~\cite{xu2023long,pramanik2025run}. A comprehensive review of active flow control using reinforcement learning has been provided by Xie et al.~\cite{xie2023deep}. Nevertheless, challenges remain in applying reinforcement learning to complex industrial processes. 

In this study, we focus on mixing, a fundamental unit operation throughout the manufacturing industry. Despite its importance, optimizing a mixing process is not necessarily straightforward. One reason is that it is often unclear which quantities should be monitored to characterize the mixing state. Another is that the flow information available for measurement in industrial settings is generally limited. If a practical framework for applying reinforcement learning to fluid mixing can be established under such constraints, its potential applications would be enormous.

\begin{figure}
\begin{center}
\includegraphics[width=0.5\textwidth]{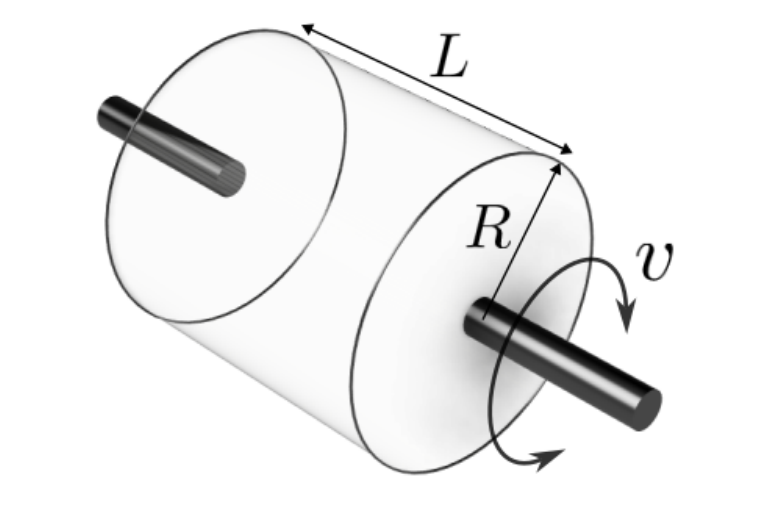}
\end{center}
\caption{\label{f:system} Schematic of the system considered. A cylindrical vessel of radius $R$ [m] and length $L$ [m] is filled with a fluid of kinematic viscosity $\nu$ [m$^2$/s]. The fluid is mixed by rotating the vessel about its axis with an angular velocity $v(t)$ [rad/s]. }
\end{figure}

In this paper, we consider fluid mixing in a system that is as simple as possible. Specifically, we examine whether a Newtonian fluid contained in a cylindrical vessel can be mixed solely by rotating the vessel about its axis (Fig.~\ref{f:system}). In this configuration, the unsteadiness of the vessel rotation is essential. A steady rotation of the vessel generates only a solid-body rotation of the fluid, resulting in extremely poor mixing. Such co-rotation is also commonly observed in stirred vessels. In conventional mixing processes, fluid co-rotation is suppressed by installing baffles in the vessel. Alternatively, varying the rotational speed or direction of an impeller in an unsteady manner can improve mixing performance by preventing co-rotation~\cite{Lamberto1996,Yao1998,Komoda2022}. However, it is highly challenging to predict how and when the rotational speed should be varied to achieve optimal mixing. Therefore, in the present study, we demonstrate the usefulness of reinforcement learning for mixing optimization by considering a system for which the timescale associated with the onset of co-rotation is theoretically known.

The optimization of fluid mixing using reinforcement learning has previously been investigated~\cite{Konishi2022,Lee2026}. However, when considering practical applications to industrial mixing processes, two challenges remain. The first is the limited availability of observable quantities. In other words, the complete state of the system required for reinforcement learning is not directly observable. The second is the issue of learning stability. In this paper, we address these challenges by exploiting insights from dynamical systems theory. For the first issue, we employ the method proposed by Kubo and Shimizu~\cite{kubo2022efficient}, which is based on the finite-dimensionality of dynamical-system attractors. For the second issue, we develop a framework for stably generating unsteady control signals by employing a dynamical system whose unique attractor is a stable limit cycle.

The optimization problem described above can be formulated as follows. Let the angular velocity of the vessel (Fig.~\ref{f:system}), $v(t)$, satisfy
\begin{equation}
 \frac{dv}{dt}
 =
 {\cal M}(\bm{s}(t);\bm{\theta})
\end{equation}
where ${\cal M}$ denotes the sum of the viscous torque and the externally applied torque. Here, $\bm{s}(t)$ denotes the state vector of the system, and $\bm{\theta}=\{\theta_0,\theta_1,\ldots\}$ represents the control parameters that determine the policy ${\cal M}$~\cite{sutton1998reinforcement}. The objective is to optimize the control parameters $\bm{\theta}$ that determine ${\cal M}$. In the present study, instead of directly observing the flow state $\bm{s}(t)$, we use the time series of the angular velocity $v(t)$. This choice is physically justified because Takens' embedding theorem~\cite{takens2006detecting} implies that the flow state can, in principle, be reconstructed from the history of the vessel angular velocity.

As a simple implementation, one may define the policy using a linear function of the form
\begin{equation}
\label{eq:mu=t0*v0+t1*v1}
{\cal M}
=
\theta_0 v(t)
+
\theta_1 v(t-1)
+
\cdots
\:.
\end{equation}
By increasing the number of delayed values included in the policy, more flexible control can be achieved. However, if, for example, $\theta_0>0$ and $\theta_i=0$ $(i=1,2,\ldots)$, the angular velocity $v(t)$ diverges, and therefore no objective function based on a time-averaged quantity can be defined. In fact, as shown in Appendix \ref{sec:fail}, reinforcement learning based on such a policy is unstable. Consequently, optimization generally requires imposing constraints on $\bm{\theta}$. This complicates the procedure. In this study, we propose an alternative approach that avoids such complexity while enabling a wide variety of motions to be generated using only a small number of parameters.

\section{Methods}

\subsection{System under consideration} 
\label{sec:system}

We investigate the following flow system. A cylindrical vessel (radius $R=0.04$ m and length $L=0.0875$ m) is filled with a Newtonian fluid (density $\rho=1000$ kg/m$^3$), and the vessel is rotated about its axis (Fig.~\ref{f:system}). When the angular velocity $v(t)$ is kept constant, the flow inside the vessel eventually settles into a solid-body rotation state (co-rotation). The relaxation toward the co-rotation state is governed by the faster of two mechanisms: viscous diffusion from the vessel wall and spin-up~\cite{watkins1977spin}. For a cylinder rotating at a constant angular velocity $v_c$ ($>0$), the Ekman number is defined as
\begin{equation} 
\label{eq:E-def} 
 E=\frac{\nu}{v_cL^2}. 
\end{equation} 
The viscous timescale is then estimated as 
\begin{equation} 
 \label{eq:Tvis-def}
 T_\text{vis} 
 \sim 
 E^{-1}(R/L)^2v_c^{-1}
 \sim 
 R^2/\nu \:, 
\end{equation} 
while the spin-up timescale is estimated as 
\begin{equation} 
 \label{eq:Tspin-def}
 T_\text{spin} 
 \sim 
 E^{-1/2}v_c^{-1}
 \sim 
 L/\sqrt{\nu v_c}\:. 
\end{equation} 
Therefore, when $T_\text{spin}<T_\text{vis}$, the flow is brought to the co-rotation state through spin-up~\cite{watkins1977spin}.

No effective mixing occurs in a solid-body rotation flow. Therefore, if the fluid is to be mixed solely by rotating the vessel, unsteady vessel rotation is essential~\cite{Aref1984}. In the following, reinforcement learning is used to determine what type of unsteady rotation can generate stronger axial flows. In the optimization, however, the magnitude of the angular velocity $v(t)$ is constrained to remain below $v_0$ ($=2\pi$ rad/s).

Two cases are considered for the kinematic viscosity of the fluid. First, in \S~\ref{sec:opt}, the viscosity is fixed at $\nu=10^{-4}$ m$^2$/s. In this case, the Reynolds number is estimated as
\begin{equation}
 Re=\frac{vR^2}{\nu}<\frac{2\pi\times 0.04^2}{10^{-4}}\simeq100
\end{equation}
Thus, only laminar flow can be sustained in the vessel. In \S~\ref{sec:zonen}, we also consider a time-dependent viscosity varying from $\nu=10^{-4}$ to $10^{-3}$ m$^2$/s. In this case, the Reynolds number becomes even smaller. Consequently, mixing is particularly challenging, and it is the axial flows generated during spin-up and spin-down that promote fluid mixing.

The flow in the vessel is obtained by numerically solving the incompressible Navier--Stokes equations,
\begin{equation}
\frac{\partial\bm{u}}{\partial t}
+
\bm{u}\cdot\bm{\nabla}\bm{u}
=
-\frac1\rho\:\bm{\nabla}p
+
\nu\nabla^2\bm{u}
\end{equation}
together with the continuity equation $\bm{\nabla}\cdot\bm{u}=0$, subject to the no-slip boundary condition on the inner wall of the vessel.

As described in Ref.~\onlinecite{watanabe2025}, the numerical method is based on a finite-difference scheme on a staggered grid. Spatial derivatives are approximated using the second-order central-difference scheme, while time integration is performed using the first-order explicit Euler method. A cubic computational domain with a side of 0.1 m is discretized using $128^3$ grid points, and the cylindrical vessel is represented by the immersed boundary method.

\subsection{Reinforcement-learning framework}

As discussed in \S~\ref{sec:intro}, when optimizing a mixing process (or, more generally, a flow), it is often difficult to accurately measure the state of the system. Therefore, rather than directly observing the flow state in the vessel, we exploit the fact that the internal state is uniquely determined by the vessel motion, i.e., the time series of the boundary condition. Specifically, the angular velocity $v$ is used as the observable quantity, and the applied torque is controlled based on it. However, if the torque applied to the vessel is represented by the simplest policy, (\ref{eq:mu=t0*v0+t1*v1}), which uses the history of the angular velocity $v(t)$, the resulting control becomes unstable (Appendix \ref{sec:fail}).

To overcome this difficulty, we incorporate insights from dynamical systems theory. Specifically, we consider the following dynamical system, consisting of the vessel angular velocity $v(t)$ and an auxiliary variable $x(t)$, where $x(t)/\omega$ represents the rotation angle of the vessel. The system is based on the lowest-order polynomial representation of the Li\'enard equation~\cite{lins2006lienard}:
\begin{equation}
\label{eq:xv-x}
\frac{dx}{dt}=\omega v 
\end{equation}
\vspace*{-20pt}
\begin{subequations}
\label{eq:xv-v}
\begin{empheq}
[left={\displaystyle\frac{dv}{dt}= \empheqlbrace \,}]{alignat=2}
& -\omega x + \omega \mu v(v_0^2-v^2) & \quad (|v| \leq v_0) \label{eq:xv-v-1}\\
& -\omega v & \quad (|v| > v_0) \label{eq:xv-v-2}
\end{empheq}
\end{subequations}
where $\omega$ ($>0$) [s$^{-1}$] is a parameter controlling the characteristic timescale, and $\mu$ ($>0$) [s$^2$] is a parameter characterizing the shape of the attractor.

Most importantly, for any given set of parameters $\{\omega,\mu\}$, the dynamical system represented by (\ref{eq:xv-x}) and (\ref{eq:xv-v}) possesses a unique stable limit cycle as its attractor. Recall that the control is performed under the constraint that the magnitude of the angular velocity $v(t)$ must remain below $v_0$ ($=2\pi$ rad/s). The branch (\ref{eq:xv-v-2}) is introduced to ensure this constraint.

\begin{figure}
\centering
\includegraphics[width=0.8\columnwidth]{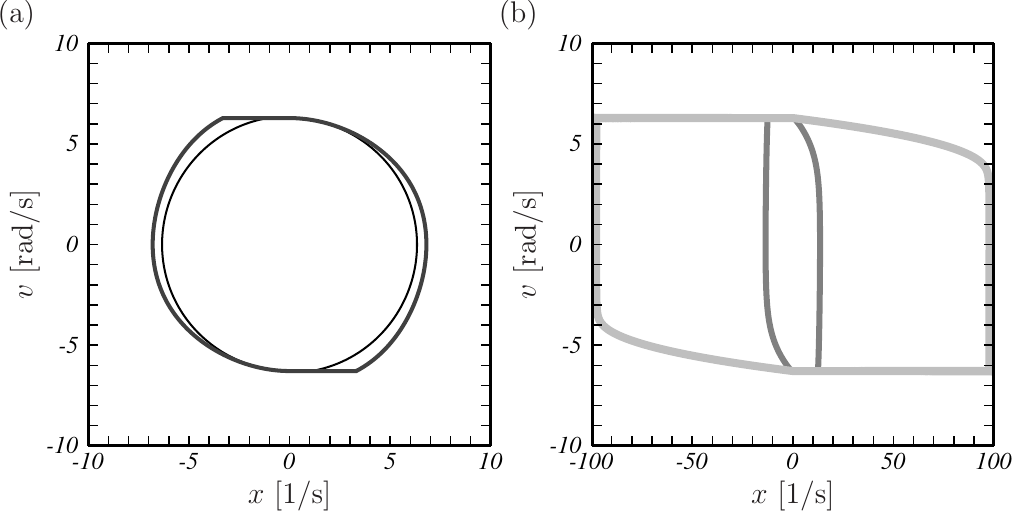}
\caption{\label{f:cycle-1}
Dependence of the limit cycle defined by (\ref{eq:xv-x}) and (\ref{eq:xv-v}) on $\mu$. (a) $\mu=0.001$ s$^2$ and $0.01$ s$^2$. (b) $\mu=0.1$ s$^2$ and $1$ s$^2$. The thicker line, shown in a lighter shade, corresponds to the larger value of $\mu$.}
\vspace*{20pt}
\centering
\includegraphics[width=0.8\columnwidth]{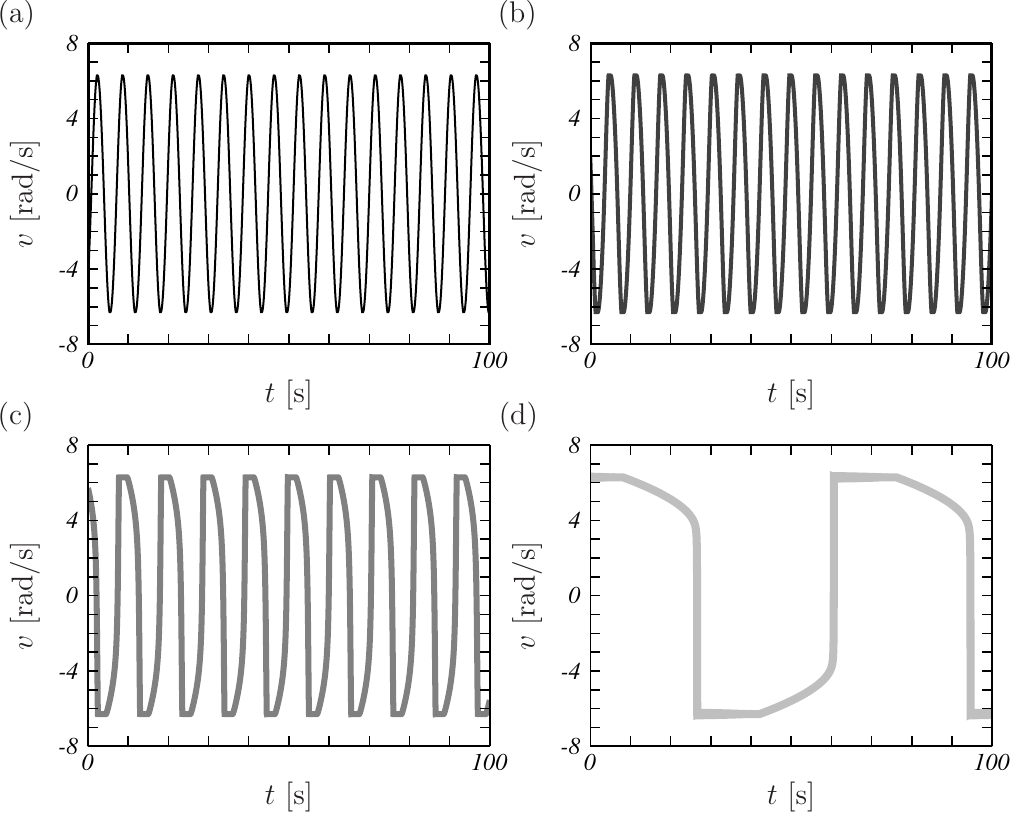}
\caption{\label{f:cycle-2} 
Time series of the angular velocity $v(t)$ for the limit cycles shown in Fig.~\ref{f:cycle-1}. (a) $\mu=0.001$ s$^2$, (b) $0.01$ s$^2$, (c) $0.1$ s$^2$, and (d) $1$ s$^2$. In all cases, $\omega=1$ s$^{-1}$. }
\end{figure}

In fact, (\ref{eq:xv-x}) and (\ref{eq:xv-v}) can represent a wide variety of limit cycles depending on the values of $\{\omega,\mu\}$. As examples, the limit cycles in the $x$--$v$ plane for $\mu=0.001$, $0.01$, $0.1$, and $1$ s$^2$ are shown in Fig.~\ref{f:cycle-1}. The shape of the limit cycle changes with $\mu$. As mentioned above, $\omega$ only modifies the timescale and therefore does not alter the limit-cycle shape. The corresponding time series for the values of $\mu$ shown in Fig.~\ref{f:cycle-1} are presented in Fig.~\ref{f:cycle-2}. It can be seen that $\mu$ controls the residence time near the maximum value of $|v|$. In fact, (\ref{eq:xv-v}) reduces to a sinusoidal function with a period of $2\pi/\omega$ when $\mu=0$. As $\mu$ increases, the time series of $v$ approaches a square wave.

Here, we emphasize the advantages of using the limit cycles represented by the dynamical system (\ref{eq:xv-x}) and (\ref{eq:xv-v}) as the angular velocity $v(t)$ of the rotating vessel. Since a unique stable periodic solution exists for any given set of parameters $\{\omega,\mu\}$, the control remains stable and the angular velocity $v(t)$ does not diverge. As a consequence, a steady-state process can be realized. This makes it possible to define the objective function as the long-time average of the reward, which is difficult to achieve using a simple policy such as (\ref{eq:mu=t0*v0+t1*v1}). Moreover, whereas a stable fixed point corresponds to a steady rotation and is therefore ineffective for mixing, the proposed approach enables policy optimization within a family of stable periodic motions.

The reinforcement learning task is therefore to optimize the limit-cycle parameters $\bm{\theta}=\{\omega,\mu\}$. The state variables are the rotation angle $x/\omega$ and the angular velocity $v$, the action is the torque $dv/dt$, and the reward $E_z$ is defined as the kinetic energy associated with the axial velocity $u_z$,
\begin{equation}
 E_z
 =
 \int_{V} \frac{1}{2}\rho u_z^2 dV    
 \:
\end{equation}
The objective function is defined as its time average,
\begin{equation}
\label{eq:barEz}
 \overline{E_z}=\lim_{T_\text{a}\to \infty} \frac{1}{T_\text{a}} \int_{0}^{T_\text{a}} E_z dt
 \:.
\end{equation}
The reinforcement learning algorithm then seeks the parameters $\{\omega,\mu\}$ that maximize $\overline{E_z}$. For this purpose, we employ the reinforcement-learning method for partially observable systems proposed by Kubo and Shimizu~\cite{kubo2022efficient}.

An episode is defined as a 20-s interval during which the search direction for the parameters $\bm{\theta}=\{\omega,\mu\}$ is fixed. During the first 10 episodes, the search direction is varied randomly. Thereafter, it is updated according to the stochastic gradient of the objective function with respect to $\bm{\theta}$~\cite{kubo2022efficient}. Appropriate tuning of hyperparameters, such as the discount factor, learning rate, and normalization of actions and states, is required. Details are available in the repository\footnote{Optimization of Mixing, https://github.com/RL-for-fluid/mixing.git}.

\section{Results} 

\subsection{Optimization of mixing} 
\label{sec:opt} 

\begin{figure}
\includegraphics[width=\columnwidth]{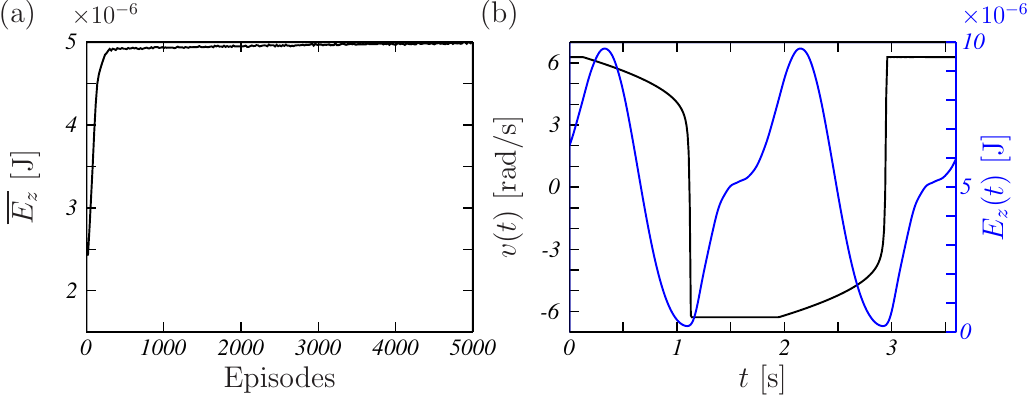}
\caption{\label{f:optimal}(a) Episode dependence of the objective function $\overline{E_z}$ when reinforcement learning is performed with the kinematic viscosity fixed at $\nu=10^{-4}$ m$^2$/s. (b) Time series of the angular velocity $v$ and the reward $E_z$ for the optimal parameters over one period.}
\end{figure}

In this section, we present the results of maximizing $\overline{E_z}$ using the reinforcement-learning framework formulated in the previous section. In this optimization, the parameters $\omega$ and $\mu$, which determine the policy ${\cal M}$ governing the periodic motion of the vessel, are optimized. Their initial values can be chosen arbitrarily; here, we set $\omega=2.72(=e)$ s$^{-1}$ and $\mu=0.368(=1/e)$ s$^2$. The optimization process obtained using the proposed method is shown in Fig.~\ref{f:optimal}(a). Stable optimization is achieved as learning progresses, and the optimization converges after approximately 5000 episodes.

The optimal control yields the parameters $\bm{\theta}^*=\{\omega^*,\mu^*\}=\{7.96~\text{s}^{-1},\,0.399~\text{s}^2\}$. The corresponding period is 3.64 s.

Meanwhile, Table~\ref{t:omega-mu} presents the results of a parameter sweep of $\overline{E_z}$ under the constraint $|v|<v_0$, where $\overline{E_z}$ is evaluated as the average over one period after a sufficiently long time has elapsed from the initial condition. The maximum value, $\overline{E_z}=4.9\times10^{-6}$ J, is obtained at $\omega=6$ s$^{-1}$ and $\mu=0.3$ s$^2$. The reinforcement-learning result shown in Fig.~\ref{f:optimal}(a) converges to a parameter region close to the optimum identified by the parameter sweep in Table~\ref{t:omega-mu}.

\begin{table}
\centering
\begin{tabular}{|c||c|c|c|c|c|}  \hline
& $\omega=0.3$ s$^{-1}$ & $\omega=1$ s$^{-1}$& $\omega=3$ s$^{-1}$& $\omega=6$ s$^{-1}$ & $\omega=10$ s$^{-1}$\\ \hline \hline
$\mu=0.001$ s$^2$& $6.4 \times 10^{-7}$ J & $3.8 \times 10^{-6}$ J & $2.9 \times 10^{-6}$ J & -- & --\\ \hline
$\mu=0.003$ s$^2$&  $6.5 \times 10^{-7}$ J & $3.9 \times 10^{-6} $ J & $3.0 \times 10^{-6}$ J & -- & -- \\ \hline
$\mu=0.01$  s$^2$&  $6.7 \times 10^{-7}$ J & $3.9 \times 10^{-6}$ J & $3.2 \times 10^{-6}$ J & $2.4 \times 10^{-6}$ J & $1.3 \times 10^{-6} $ J \\ \hline
$\mu=0.03$  s$^2$& $6.6 \times 10^{-7}$ J & $3.6 \times 10^{-6}$ J & $3.5 \times 10^{-6}$ J & $3.0 \times 10^{-6} J $ & $1.7 \times 10^{-6} $ J \\ \hline
$\mu=0.1$   s$^2$& $5.0 \times 10^{-7}$ J & $2.2 \times 10^{-6}$ J & $4.7 \times 10^{-6}$ J & $4.1 \times 10^{-6}$ J & $3.4 \times 10^{-6}$ J \\ \hline  
$\mu=0.3$   s$^2$& -- & --  & $3.3 \times 10^{-6}$ J & $4.9 \times 10^{-6}$ J & $4.3 \times 10^{-6}$ J \\ \hline  
$\mu=1$     s$^2$& -- & --  & $9.8 \times 10^{-7}$ J & $2.3 \times 10^{-6}$ J & $ 3.9 \times 10^{-6}$ J \\ \hline  
\end{tabular}
\caption{
\label{t:omega-mu}
Results of a parameter sweep of $\overline{E_z}$ [J] over the control parameters $\omega$ and $\mu$. Here, $\overline{E_z}$ is evaluated as the average over one period after a sufficiently long time has elapsed from the initial condition.}
\end{table}

The vessel motion obtained by reinforcement learning is also physically reasonable. Figure~\ref{f:optimal}(b) shows the time series of the angular velocity $v(t)$ and $E_z(t)$ for the optimal control in the final episode. It can be seen that, under the optimal control, the reversal of the rotation begins when $E_z(t)$ approaches zero. In other words, when continuous rotation in one direction causes the fluid to approach a co-rotation state, the vessel reverses its rotation before co-rotation is established.

We further verify that the period obtained by reinforcement learning is theoretically reasonable. As discussed in \S~\ref{sec:system} and described in Ref.~\onlinecite{watkins1977spin}, the timescale over which the fluid approaches co-rotation is determined by the shorter of the spin-up time $T_\text{spin}$, estimated by (\ref{eq:Tspin-def}), and the viscous time $T_\text{vis}$, estimated by (\ref{eq:Tvis-def}). For $\nu=10^{-4}$ m$^2$/s, $L=0.0875$ m, $R=0.04$ m, and $v_c=v_0=2\pi$ rad/s, $T_\text{spin}\approx 3.4$ s and $T_\text{vis}\approx 16$ s. Therefore, under the present conditions, the co-rotation timescale is determined by $T_\text{spin}$. The period of the vessel motion corresponding to the optimal parameters is 3.64 s, which is of the same order as $T_\text{spin}$. Moreover, Fig.~\ref{f:optimal}(b) directly confirms that the optimal control effectively prevents co-rotation.

Recall that the waveform changes from sinusoidal to square-wave-like as $\mu$ increases, thereby changing the residence time near the maximum value of $|v(t)|$. Although it is difficult to predict a priori what residence time is optimal, the vessel motion optimized by the reinforcement-learning method is closer to a square wave than to a sinusoidal function.

\begin{figure}
\includegraphics[width=0.85\textwidth]{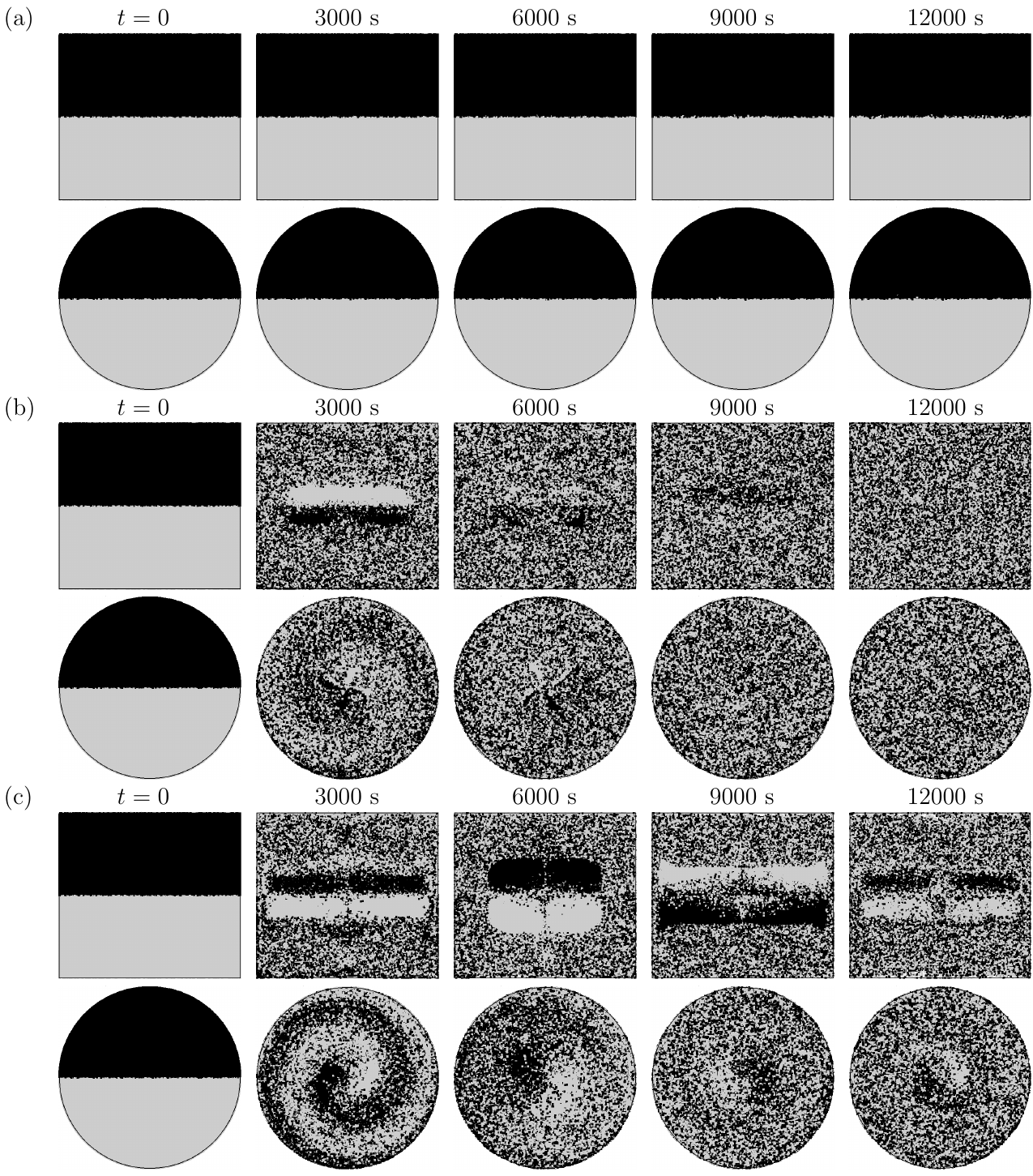}
\caption{\label{f:mixing}
Mixing under (a) solid-body rotation, (b) the optimal parameters $\{\omega^*,\mu^*\}=\{7.96~\text{s}^{-1},\,0.399~\text{s}^2\}$, for which the corresponding period of the vessel motion is $T=3.64$ s, and (c) the initial parameters $\{\omega,\mu\}=\{2.72~\text{s}^{-1},0.368~\text{s}^2\}$ without optimization. Mixing of two fluids, initially separated by a plane passing through the cylinder axis, is visualized by fluid particles whose trajectories are tracked by numerically integrating the advection equation (\ref{eq:advection}). Snapshots are shown at $t=0$, $3000$, $6000$, $9000$, and $12000$ s.}
\end{figure}

Although the optimization maximizes the objective function $\overline{E_z}$ based on the axial velocity $u_z$, it is also important to confirm that mixing is indeed enhanced. Figure~\ref{f:mixing} shows the mixing process of two fluids that are initially separated by a plane passing through the cylinder axis. Approximately $4.4\times10^5$ fluid particles are placed in each region, and their trajectories $\bm{x}_p(t)$ are numerically tracked by integrating the advection equation,
\begin{equation}
\label{eq:advection}
 \frac{d\bm{x}_p}{dt}
 =
 \bm{u}(\bm{x}_p(t),t)
\end{equation}
using a second-order Runge--Kutta method. The particle velocity $\bm{u}(\bm{x}_p(t),t)$ is evaluated by linearly interpolating the velocity field defined on the grid points.

It can be seen that no mixing occurs in the solid-body rotation flow shown in (a), i.e., the co-rotation state. In contrast, under the optimal parameters shown in (b), substantial mixing is achieved within approximately 6000 s. Panel (c) shows the result for the initial parameters $\{\omega,\mu\}=\{2.72~\text{s}^{-1},\,0.368~\text{s}^2\}$ without any optimization. Compared with (c), mixing in (b) proceeds significantly faster. Thus, we can confirm that fluid mixing is indeed optimized by the reinforcement-learning method.

\begin{figure}
\begin{center}
\includegraphics[width=0.5\columnwidth]{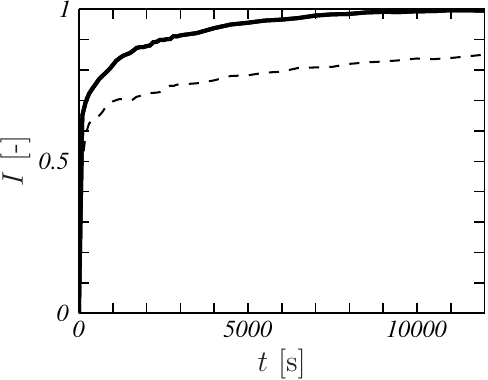} 
\end{center}
\caption{\label{f:mixing-index} 
Time evolution of the mixing index $I$ for the two cases shown in Fig.~\ref{f:mixing}(b) and (c). The solid and dashed lines correspond to the cases with and without optimization, respectively.}
\end{figure}

To quantify the difference observed in Fig.~\ref{f:mixing}, the mixing index proposed in Refs.~\onlinecite{Goto2014,Watanabe2022} is shown in Fig.~\ref{f:mixing-index}. Specifically, the fluid domain is divided into cubic cells of size $(R/12)^3$. Let $p$ denote the fraction of one of the two fluids within each cell. Since the standard deviation $\sigma$ of $p$ is zero for a perfectly mixed state and $1/2$ for a completely segregated state, the mixing index is defined as 
\begin{equation} 
I=1-2\sigma\:. 
\end{equation} 
Figure~\ref{f:mixing-index} quantitatively confirms the improvement in mixing achieved with the optimized parameters.

\subsection{Optimization with gradually increasing viscosity}
\label{sec:zonen}

In the previous subsection, we considered the case in which the fluid properties remain unchanged. In practical applications, however, the fluid properties often change as mixing proceeds. A typical example is the addition of a thickening agent. In this subsection, we consider the case in which the fluid viscosity varies with time as a result of mixing, and show that the proposed reinforcement-learning method works even under such conditions.

Specifically, we consider the case in which the kinematic viscosity of the fluid varies with time according to
\begin{equation}
\label{eq:nu-t-dep}
 \nu(t)
 =
 \nu_0
 +
 \dot{\nu}\:t
 \:,
\end{equation}
where $\nu_0=10^{-4}$ m$^2$/s and $\dot{\nu}=2.25\times10^{-8}$ m$^2$/s$^2$. This corresponds to a linear increase of the viscosity from $\nu=10^{-4}$ to $10^{-3}$ m$^2$/s over 40000 s. The value of $\nu_0$ is identical to that used in the optimization presented in the previous subsection.

As discussed in the previous subsection, when $\nu=10^{-4}$ m$^2$/s, the flow approaches a co-rotation state through the spin-up process. Since the spin-up time depends on the kinematic viscosity of the fluid, the optimal parameters vary over time. In the following, we show that the proposed method can successfully identify this variation. 

\begin{figure}
\includegraphics[width=\textwidth]{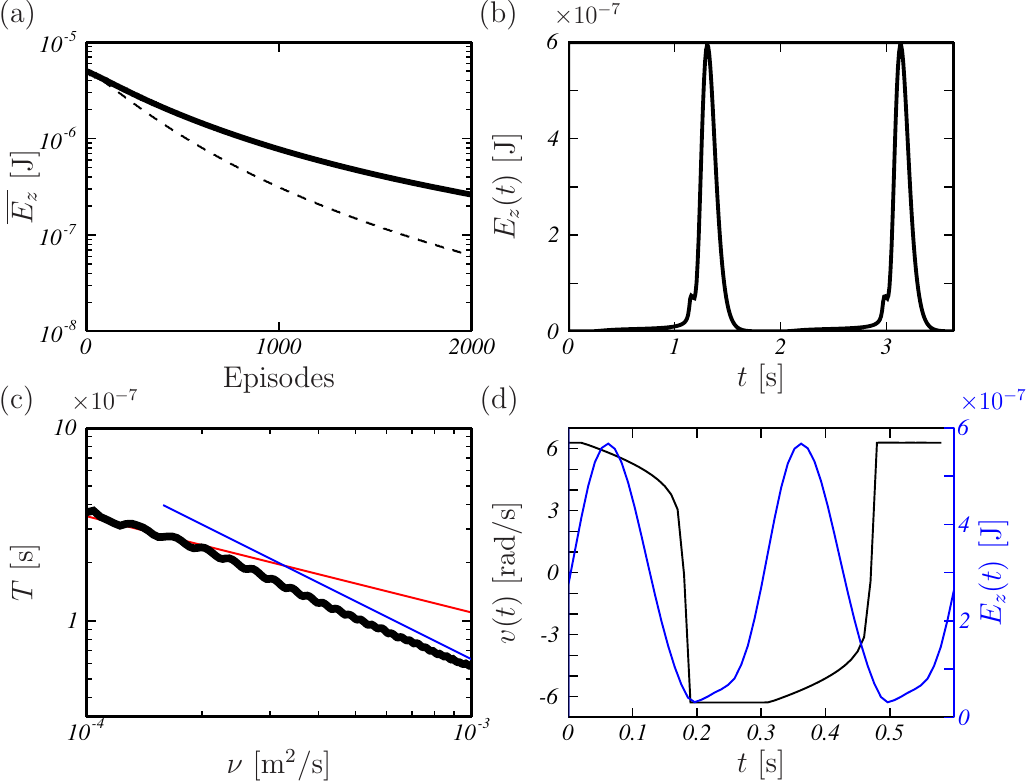}
\caption{\label{f:t-dep}
Results of the optimization for the time-dependent viscosity given by (\ref{eq:nu-t-dep}). (a) Episode dependence of the objective function. The solid line shows the result obtained using the proposed reinforcement-learning method. The dashed line shows the result obtained without optimization, where the control parameters optimized for $\nu=\nu_0=10^{-4}$ m$^2$/s are not updated. (b) Time dependence of the reward for a fluid with $\nu=10^{-3}$ m$^2$/s using the control parameters optimized for $\nu=10^{-4}$ m$^2$/s. (c) Dependence of the optimized period $T$ of the vessel motion on the kinematic viscosity $\nu$. The red line indicates the spin-up scaling, and the blue line indicates the viscous-timescale scaling. (d) Time series of the vessel angular velocity $v$ and the reward for the optimal control parameters at the final time with $\nu=10^{-3}$ m$^2$/s.}
\end{figure}

Figure~\ref{f:t-dep}(a) shows the episode dependence of the objective function. The solid line shows the result obtained using the proposed optimization method, whereas the dashed line shows the result obtained without optimization, i.e., when the parameters are fixed at the values obtained in the previous subsection, $\{\omega,\mu\}=\{7.96~\text{s}^{-1},\,0.399~\text{s}^2\}$. Although $\overline{E_z}$ decreases even with optimization because increasing viscosity makes it more difficult to generate axial flow, substantially larger values are maintained than in the case without optimization. Indeed, Fig.~\ref{f:t-dep}(b) shows that the periods during which the reward remains close to zero become longer when no optimization is performed. Since the increase in viscosity shortens the spin-up time, the vessel motion must be re-optimized to avoid wasting time.

The solid line in Fig.~\ref{f:t-dep}(c) shows the period $T$ of the vessel motion at the optimal parameters as a function of the kinematic viscosity. Since the spin-up time $T_\text{spin}$ decreases as the viscosity increases, the reinforcement-learning method successfully adapts the vessel motion accordingly. The red line in this figure represents the viscosity dependence of the spin-up time $T_\text{spin}$ (see (\ref{eq:Tspin-def})). Although the reinforcement-learning method proposes periods shorter than this scaling, this deviation reflects the competing effect of the viscous timescale $T_\text{vis}$, whose scaling is shown by the blue line. The physical origin of this behavior is discussed in the following paragraph.

According to Ref.~\onlinecite{watkins1977spin}, the flow in a rotating cylinder is characterized by $\alpha=E^{1/2}(L/R)^2$, and viscous effects dominate when $\alpha>1$. For the viscosity variation given by (\ref{eq:nu-t-dep}), $\alpha$ increases from 0.21 to 0.68 and therefore gradually approaches the viscous-dominated regime. In fact, at the final time, the viscous timescale $T_\text{vis}$ is approximately 1.6 s, which is comparable to the spin-up time $T_\text{spin}\approx 1.1$ s for $\nu=10^{-3}$ m$^2$/s. Consequently, the flow approaches solid-body rotation through the combined effects of spin-up and viscous diffusion. This explains the transition of the optimal period $T$ from the spin-up scaling toward the viscous-timescale scaling shown in Fig.~\ref{f:t-dep}(c). The reinforcement-learning method successfully captures this qualitative change in the co-rotation timescale. Indeed, the optimal control at the final time, shown in Fig.~\ref{f:t-dep}(d), successfully prevents co-rotation.

\section{Conclusion} 

We have considered the reinforcement-learning optimization of fluid mixing. In many industrial mixing processes, unsteady motions are often essential for efficient mixing. However, determining the optimal time dependence of such motions remains a challenging problem. As a representative example, we have investigated fluid mixing in a rotating cylindrical vessel (Fig.~\ref{f:system}) and optimized the time series of the vessel angular velocity using reinforcement learning.

A major challenge in applying reinforcement learning to flow control is that the flow field is inherently high-dimensional. As a result, it is difficult to represent the state of the system accurately. This difficulty is particularly severe in practical applications, where only partial observations of the flow are available. The first key idea of the present study is to overcome this limitation by using time series of the boundary condition as the state variable. However, since a policy is generally a function of the state, the resulting control may become unstable if no constraints are imposed on its parameters (Appendix \ref{sec:fail}, Fig.~\ref{f:div}). When the objective function is defined as a long-time average, as in (\ref{eq:barEz}), such instability hinders reinforcement-learning optimization. The second key idea of the present study is therefore to constrain the policy so that the optimization remains stable.

In particular, we have focused on stable limit cycles in dissipative dynamical systems. Since fluid flows are typical dissipative systems, limit cycles have been used for the analysis of high-dimensional flow dynamics~\cite{taira2026reducing} and for the control of resonances between vortices and wing vibrations~\cite{sumanasiri2025phase}. Rather than directly treating periodic motions of the flow itself, the present study employs a dynamical system whose convergence to a limit cycle is guaranteed, thereby providing a stable policy representation for reinforcement learning.

The dynamical system defined by (\ref{eq:xv-x}) and (\ref{eq:xv-v}) contains only two parameters, $\{\omega,\mu\}$. The parameter $\omega$ determines the timescale of the policy, whereas $\mu$ controls its waveform. As $\mu$ increases, the waveform changes from sinusoidal to square-wave-like. Reinforcement learning optimizes these parameters so as to maximize the objective function, which is defined in terms of the energy of the velocity component relevant to mixing. As expected, the optimization proceeds stably (Fig.~\ref{f:optimal}(a)), yielding the optimal parameters. The resulting optimal operation (Fig.~\ref{f:optimal}(b)) is physically reasonable from the viewpoint of the characteristic timescale of the flow, maximizing axial flow while preventing co-rotation. Although the optimal waveform itself cannot be predicted from physical considerations alone, it is consistent with the results of the parameter sweep (Table~\ref{t:omega-mu}).

It should also be emphasized that the dynamical system used in the present study, (\ref{eq:xv-x}) and (\ref{eq:xv-v}), contains only two control parameters, $\bm{\theta}=\{\omega,\mu\}$. More complex limit cycles with richer temporal dynamics can be represented by increasing the number of parameters. More generally, the relationship between an observable quantity $x$ and a control variable $v$ can be described in the vicinity of a limit cycle by 
\begin{equation}
\frac{dx}{dt}=F_x(x,v;\bm{\theta}), \qquad \frac{dv}{dt}=F_v(x,v;\bm{\theta})\:. 
\end{equation}
Therefore, the proposed framework can be generalized by optimizing the control parameters $\bm{\theta}$ under constraints that ensure $(x,v)$ remain on a limit cycle. The proposed method is thus not limited to the Li\'enard system.

In addition, we have considered a reward based on the axial velocity, which is relatively easy to observe in practice, and obtained operating conditions that enhance mixing (Fig.~\ref{f:mixing}). However, the proposed framework is not restricted to this choice of objective function. For example, the objective function can readily be redefined in terms of energy efficiency. We have also performed numerical experiments that mimic a thickening process, which is common in industrial mixing applications. The results have shown that the proposed method successfully identifies the optimal parameters at each instant of time (Fig.~\ref{f:t-dep}). Since reinforcement-learning optimization can be performed stably without directly observing the detailed state of the flow, the proposed framework provides a promising approach for industrial mixing applications.

\begin{acknowledgments}
This study was supported by the JSPS Grants-in-Aid for Scientific Research (24K07323, 25K01158, 25KJ0231, 26K17306, 26K17694) and Transformative Research Areas (JP26H00389). The numerical computations were conducted using Plasma Simulator Sousei under the NIFS Collaboration Research Programs (NIFS25KISC016, NIFS26KISC034) and RIKEN Wisteria/BDEC-01 Odyssey under the HPCI System Research Project (hp260065).
\end{acknowledgments}


%

\appendix

\section{An example demonstrating instability without constraints on the policy}
\label{sec:fail}

\begin{figure}
\begin{center}
\includegraphics[width=0.5\textwidth]{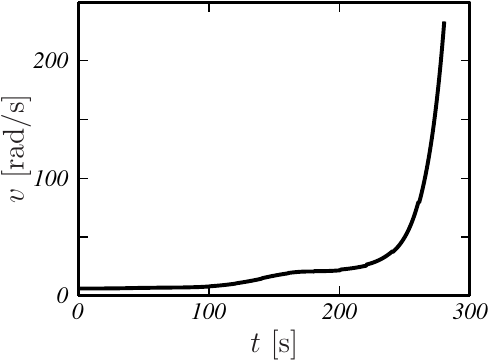} 
\end{center}
\caption{Example in which reinforcement learning becomes unstable and the angular velocity $v(t)$ diverges during training when the policy is given by (\ref{eq:ap:M=t0v0+t1v1}).}
\label{f:div}
\end{figure}

As discussed in the main text (\S~\ref{sec:intro}), the simplest control strategy is to determine the torque applied to the vessel based on the history of the angular velocity $v(t)$. Specifically, the policy ${\cal M}$ is represented by (\ref{eq:mu=t0*v0+t1*v1}), and $\theta_i$ ($i=0,1,\ldots$) are treated as control parameters. However, in such an approach, the optimization becomes unstable unless some constraints are imposed on the policy. As a specific example, suppose that the policy is represented by 
\begin{equation} 
\label{eq:ap:M=t0v0+t1v1}
{\cal M}=\theta_0 v(t)+\theta_1 v(t-1), 
\end{equation} 
using the angular velocity at two time instants. Depending on the values of $\theta_0$ and $\theta_1$, the angular velocity $v$ may diverge (Fig.~\ref{f:div}).

When the objective function is defined as the time average of the reward, a divergence of the reward during the learning process prevents further experience from being accumulated, and the learning process cannot proceed. Note that, in the example shown in Fig.~\ref{f:div}, all reinforcement-learning settings are identical to those described in the main text, except that the policy is given by (\ref{eq:ap:M=t0v0+t1v1}).
\end{document}